\documentclass{IEEEtran}

\usepackage{amsmath} 
\usepackage{amssymb} 
\usepackage{amsthm} 
\usepackage{bbm}
\usepackage{mathrsfs}
\usepackage{amssymb}
\usepackage{dsfont}
\usepackage{mathrsfs}

\usepackage{graphicx,psfrag, color}
\graphicspath{Figures/}
\usepackage{tikz}

\newtheorem{corollary}{Corollary}
\newtheorem{theorem}{Theorem}
\newtheorem{lemma}{Lemma}

\newtheorem{proposition}{Proposition}
\newtheorem{remark}{Remark}

    \newcommand{\subscr}[2]{#1_{\textup{#2}}}
    
    \newcommand{\setdef}[2]{\{#1 \, : \; #2\}}

\newcommand{\real}{\mathbb{R}}

\newcommand{\integernonnegative}{\mathbb{Z}_{\ge 0}}

\newcommand{\R}{\mathbb{R}} 
\newcommand{\N}{\mathbb{N}}  

\newcommand{\G}{\mathcal{G}} 
\newcommand{\V}{\mathcal{V}} 
\newcommand{\E}{\mathcal{E}} 

\newcommand{\neigh}{ \mathcal{N}} 	
\newcommand{\card}[1]{|#1|}  	

\def\Exp{\mathbb{E}}
\def\Prob{\mathbb{P}}

\newcommand{\pgr}{{\subscr{x}{pgr}^{\star}}} 
\newcommand{\loc}{{\subscr{x}{loc}^{\star}}} 
\newcommand{\op}{{\subscr{x}{opd}^{\star}}} 

\newcommand{\1}{\mathbf{1}} 

\newcommand{\diag}{\operatorname{diag}} 

\newcommand{\eps}{\varepsilon} 

\title{Ergodic Randomized Algorithms and Dynamics over Networks}%
\author{Chiara~Ravazzi\thanks{Chiara Ravazzi is with the Department of Electronics and Telecommunications (DET),
        Politecnico di Torino, Italy. E-mail: {chiara.ravazzi@polito.it}}, \and Paolo~Frasca\thanks{Paolo Frasca is with the Department of Applied Mathematics, University of Twente, The Netherlands. E-mail: {p.frasca@utwente.nl}},
\and Roberto~Tempo\thanks{Roberto~Tempo is with CNR-IEIIT, Politecnico di Torino, Italy. E-mail: roberto.tempo@polito.it} and
\and Hideaki~Ishii\thanks{Hideaki~Ishii is with the Department of Computational Intelligence and Systems Science,        Tokyo Institute of Technology, Japan. E-mail:        {ishii@dis.titech.ac.jp}} 
}

\date{\today}

\begin{document}

\maketitle
\begin{abstract}
Algorithms and dynamics over networks often involve randomization, and 
randomization may result in oscillating dynamics which fail to converge in a deterministic sense. In this paper, we observe this undesired feature in three applications, in which the  dynamics is the randomized asynchronous counterpart of a well-behaved synchronous one.
These three applications are network localization, PageRank computation, and opinion dynamics.
Motivated by their formal similarity, we show the following general fact, under the assumptions of independence across time and linearities of the updates: if the expected dynamics is stable and converges to the same limit of the original synchronous dynamics, then the oscillations are ergodic and the desired limit can be locally recovered via time-averaging.
\end{abstract}

\section{Introduction}

Randomization has proved to be a useful ingredient for effective algorithms in control and optimization, as reviewed in~\cite{RT-GC-FD:13}.
In network dynamics, randomization is specially natural, either by the uncertain nature of the network at hand, or by design aimed at improving performance and robustness.

In this work, we focus on a class of randomized affine dynamics which do not possess equilibria, but are stable on average. This stability property ensures that the dynamics, although it features persistent random oscillations, has an ergodic behavior. 
This ergodicity result, which we prove by classical facts of probability theory, can be readily applied to several network-based algorithms, in which randomization apparently prevents convergence.
As a consequence, the desired convergence property --holding in expectation-- can be recovered by each node through a process of time-averaging, which can be performed locally and, in some cases, without even access to a common clock.

In the examples considered in this paper, nodes interact in randomly chosen pairs, following a ``gossip'' approach which has been popularized in the field of control by~\cite{SB-AG-BP-DS:06} and has been followed in several papers. 
There has been a wide range of applications since then. Indeed, many network algorithms can be randomized in such a way that the randomized dynamics converges (almost surely) to the same limit of the synchronous dynamics.
Notable examples include consensus algorithms, studied in many papers as~\cite{FF-SZ:08a,ATS-AJ:08}, and other algorithms for estimation and classification~\cite{AC-FF-LS-SZ:10} and for optimal deployment of robotic networks~\cite{FB-RC-PF:08u}.
Nevertheless, examples of randomized algorithms which do not converge also have recently appeared in the literature. Such algorithms require some sort of additional ``smoothing'' operation in order to converge: in our approach, this goal is achieved by time-averaging.

A prime example we consider involves the problem of distributed estimation from relative measurements, which has applications from self-localization in robotic networks to synchronization in networks of clocks and to phase estimation in power grids. This problem was first introduced in a least-squares formulation in the context of clock synchronization~\cite{AG-PRK:06a} and then studied in much detail in \cite{PB-JPH:07,PB-JPH:08,PB-JPH:09,SB-SDF-LS-DV:10,WSR-PF-FF:12}, where both fundamental performance limitations and distributed algorithms have been presented. More recently, randomized algorithms for its solution have been proposed by several researchers \cite{NMF-AZ:12,RC-AC-LS-MT:13}. Regarding this problem, our contribution includes a randomized asynchronous algorithm, in which nodes update in pairs in a {\em gossip} fashion: its novelty is further discussed in Section~\ref{sect:random-localization}. A related but different randomized algorithm for least-squares estimation has been recently proposed in~\cite{KY-SS-LQ:14}.

A second example is PageRank computation, which has attracted much attention in recent years for the importance of its applications~\cite{SB-LP:98,KB-TL:06,ANL-CDM:06} and for its similarities with the consensus problem, as illustrated in~\cite{HI-RT:10}. Randomized algorithms for PageRank computation have been studied in a series of papers, including~\cite{HI-RT:10,Nazin:2011:RAD:1960870.1960904,HI-RT-EWB:12b,DBLP:journals/corr/abs-1305-3178}.
Other recent references on PageRank are listed in \cite{journals/corr/abs-1203-6606,fercoq:hal-00782749, Blondel13}.
Our contribution provides a general convergence result for randomized algorithms, which we apply to a novel pair-wise {\it gossip} algorithm in Section~\ref{sect:random-pagerank}.

A third example comes from social sciences, where there has been long-time interest in the mechanisms of opinion evolutions. It comes out that opinion dynamics models, where agents have some degree of obstinacy and interactions are randomized, gives rise to ergodic oscillations. This observation has first been made in~\cite{DA-GC-FF-AO:11} and here we extend it in connection with the Friedkin and Johnsen's model~\cite{NEF-ECJ:99} from social sciences. We propose in Section~\ref{sect:random-friedkin} a {\it gossip} mechanism of update for the opinions, which allows us to interpret the classical opinion dynamics --which makes simplistic assumptions on the communication process among individuals-- as the ``average'' evolution of our randomized model. This observation answers an open question on modeling the communication process which was raised in the original paper~\cite{NEF-ECJ:99}.

Preliminary versions of part of our results have been reported in the proceedings of technical conferences as~\cite{CR-PF-HI-RT:13a,CR-PF-RT-HI:13b}, regarding relative localization, and as~\cite{PF-CR-RT-HI:13c}, regarding opinion dynamics. 
The current presentation incorporates and builds upon the previous ones. Additionally, it includes the case study of PageRank computation, and most importantly embeds them into a comprehensive framework which is suitable for the study of other applications.

\subsection{Paper outline} In Section~\ref{sect:synchro} we study deterministic synchronous dynamics, presenting the three examples of relative localization, PageRank computation, and opinion dynamics. These dynamics are then translated into corresponding randomized asynchronous dynamics in Section~\ref{sect:randomized}. Additional remarks and research outlooks are given in a concise Section~\ref{sect:conclusion}, and an Appendix contains the technical derivation of our main result.

\subsection{Notation and preliminaries}
Throughout this paper, we use the following notation. 
Real and nonnegative integer numbers are denoted by $\real$ and $\integernonnegative$, respectively. 
The symbol $\card{\cdot}$ denotes either the cardinality of a set or the absolute value of a real number.
The symbol $e_i$ is the vector with the $i$-th entry equal to 1 and all the other elements equal to $0$, and we write $\1$ for the vector with all entries equal to 1. A vector $x$ is stochastic if its entries are nonnegative and $\sum_ix_i=1$.
A matrix $A$ is row-stochastic (column-stochastic) when its entries are nonnegative and $M\1=\1$ ($M^{\top}\1=\1$). A matrix is doubly stochastic when it is both row and column-stochastic. A matrix $P$ is said to be Schur stable if the absolute value of all its eigenvalues is smaller than 1. 
A graph is a pair $\mathcal{G}=(\mathcal{V, E})$, where $\V$ is the set of nodes and $\mathcal E\subseteq \mathcal{V\times V}$ is the set of edges. 
To avoid trivialities, we implicitly assume that graphs have at least three nodes, i.e., $\card{\V}>2$.
A graph $\mathcal{G}$ is called strongly connected if there is a path from each vertex in the graph to every other vertex.
To any matrix $P\in\R^{\mathcal V\times\mathcal V}$ with non-negative entries, we can associate a graph $\mathcal G_P=(\mathcal V, \mathcal{E}_P)$ by putting $(i,j)\in \mathcal{E}_P$ if and only if $P_{ij}>0$. The matrix $P$ is said to be adapted to graph $\mathcal G$ if $\mathcal G_P\subseteq \mathcal G$.

\section{Synchronous affine dynamics over networks}\label{sect:synchro}
Consider the affine dynamics representing  a time-invariant discrete-time dynamical system with state $x(k) \in\real^\V $, ${k\in\integernonnegative}$
\begin{equation}
\label{eq:synchro-algo}
x(k+1) = P x(k) + u
\end{equation}
with $P\in\real^{\V\times\V} $ and constant input $u\in\mathbb{R}^{\V}$.

%
%

\begin{proposition}
\label{prop:synch-converge}
If $P$ is Schur stable, then the dynamics in~\eqref{eq:synchro-algo} converges to 
$$
x^{\star}=(I-P)^{-1}u
$$
for any initial conditions $ x(0)=x_0$.
\end{proposition}
\begin{IEEEproof}
Equation~\eqref{eq:synchro-algo} implies that
$ x(k)=P^k  x(0)+\sum_{\ell=0}^{k-1} P^\ell u.$ Since all the eigenvalues of $P$ lie in the open unit disk, we have
$$
\lim_{k\to \infty}P^k=0
$$  
and 
$$
\lim_{k\to \infty}\sum_{\ell=1}^{k-1} P^\ell y=  (I-P)^{-1} u.
$$ 
These limits imply that $x^\star$ is the convergence value. 
\end{IEEEproof}

More specifically, in this paper we study affine dynamics over a certain network, described by a graph $\G=(\V,\E)$ with $n$ nodes, that is, such that {\em the matrix $P$ is adapted to the graph $\mathcal{G}$.}
%
%
%
In the rest of this section, we review the three applications of affine dynamics over networks which we have mentioned in the introduction. Even though these applications are quite diverse, we show that the algorithms for their solutions can be represented by the affine dynamics~\eqref{eq:synchro-algo}, provided suitable manipulations are performed.

\subsection{Sensor localization in wireless networks}
\label{es1}
In sensor localization in wireless networks, we seek to estimate the relative position of sensors using noisy relative measurements. We formulate the problem using an oriented graph\footnote{
An oriented graph $\G=(\V,\E)$ is a graph such that $(i,j)\in \E$ only if $i<j$. $\G$ is said to be weakly connected if the graph $\G'=(\V,\E')$ where $\E'=\setdef{(h,k)\in \V\times \V}{\text{either $(h,k)\in \E$ or $(k,h)\in \E$}}$  has a path which connects every pair of nodes.} 
 $\mathcal{G}=(\V, \E)$.
Each node $i$ in $\V$ has to estimate its own variable 
$s_i$, knowing only noisy measurements of some difference with neighboring edges
$$
b_{i,j}= s_i- s_j+\eta_{i,j} \qquad \text{if $(i,j)\in \E$ or $(j,i)\in \E$}
$$
where $\eta_{i,j}$ is additive noise.
The graph topology is encoded in the incidence matrix $A\in\{0,\pm1\}^{\mathcal{E}\times\V}$ defined by 
$$
A_{ei}=\begin{cases}
+1&\text{if } e= (i,j)\\
-1&\text{if } e=(j,i)\\
0&\text{otherwise}\\
\end{cases}
$$ for every $e\in \mathcal{E}$. We can collect all the measurements and variables in vectors $b\in \mathbb{R}^{\mathcal{E}}$ and $s\in \real^\V$, so that 
$$ b=A s +\eta,
$$
where $\eta\in\mathbb{R}^{\mathcal{E}}$. 
A least-squares approach can be used to determine the best estimate of the state $s$ based on the measurements $b$. That is, we define the unconstrained quadratic optimization problem
\begin{equation}
\label{LS}
\min_{z} \|A z -b\|_2^2
\end{equation}
where $\|\cdot\|_2$ denotes the Euclidean norm.
The solution of this problem is summarized in the following standard result.
\begin{lemma}[Least-squares localization]
\label{lemma:centralized-LS} 
Given a weakly connected oriented graph
$\mathcal{G}$ with incidence matrix $A$, let $\mathcal{X}$ be the set of solutions of \eqref{LS} and let $L:=A^\top A$ be the Laplacian. The following facts hold:
\begin{enumerate}
\item $x \in \mathcal{X}$ if and only if $A^\top Ax =A^\top b$;
\item there exists a unique minimizer $ \loc \in \mathcal{X}$ such that $\| \loc \|_2=\min_{z\in \mathcal{X}}\| z\|_2$;
\item $ \loc =L^{\dag} A^\top b$,
where $L^{\dag}$ is the pseudoinverse matrix.
\end{enumerate}
\end{lemma}

We note that the optimal least-squares solution $ \loc $ is the minimum-norm element of the affine space of solutions of~\eqref{LS} and $A\loc=A( \loc +c\1)$ for any scalar~$c$.
As shown in Lemma~\ref{lemma:centralized-LS}, the solution to the least-squares relative localization problem \eqref{LS} is explicitly known. Furthermore, it can be easily computed by an iterative gradient algorithm.
Given a parameter $\tau>0$ and the initial condition $x(0)=0$, we let 
\begin{align}\label{grad_des}
x(k+1)=(I-\tau L)x(k)+\tau A^\top b
\end{align}
where the matrix $I-\tau L$ is doubly stochastic. For this reason, it holds true that
\begin{align*}
\1^{\top}x(k+1)&=\1^{\top}(I-\tau L)x(k)+ \tau\1^{\top} A^\top b
=\1^{\top}x(k)
\end{align*}
for all $k\in\integernonnegative$.
Defining $\Omega=I-\1\1^{\top}/n$, we have that $\Omega x(k)=x(k+1)- \frac1n \1^{\top}x(k) \1=x(k)$
because $x(0)=0$ and moreover
\begin{align*}
\Omega x(k+1)&=(I-\tau L)\Omega x(k)+ \tau A^\top b
\end{align*}
because $\Omega (I-\tau L)=(I-\tau L)\Omega$ and $\Omega A^\top b=A^\top b.$
Then, equation~\eqref{grad_des} can be rewritten as
$$ x(k+1)=(I-\tau L)\Omega x(k)+ \tau A^\top b $$
and falls in the class of affine dynamics
(\ref{eq:synchro-algo}) taking
\begin{equation}\label{Pu-def-localization} P=(I-\tau L)\Omega \quad \text{and} \quad u=\tau A^\top b.\end{equation}
The convergence properties of this algorithm are summarized in the following simple result (also available as~\cite[Proposition~1]{WSR-PF-FF:12}), 
where we let $d_{\max}$ be the maximum degree given by $d_{\max}=\max_i\card{\{(i,j)\in \E\}\cup \{(j,i)\in \E\}}$.
\begin{proposition}[Convergence of gradient descent algorithm]
Assume that the graph $\mathcal{G}$ is weakly connected. Then, the gradient descent algorithm in~\eqref{grad_des} with $x(0)=0$ converges to the optimal least-squares solution $ \loc $ if $\tau<1/d_{\max}$. 
\end{proposition}

\begin{IEEEproof}
By Gershgorin Circle Theorem and~\cite[Theorem~1.37]{FB-JC-SM:09}, the eigenvalues of $L$ satisfy $0=\lambda_1<\lambda_i\le\lambda_{i+1}\le \lambda_n\le 2{d}_{\max}$. 
Then, it is clear that the spectral radius of $(I-\tau L)\Omega$ is equal to $\max\{1-\tau \lambda_1,-1+\tau\lambda_{n}\}$. 
By the assumption on $\tau$, we thus conclude that $(I-\tau L)\Omega$ is Schur stable, and convergence to the least-squares solution immediately follows by Proposition~\ref{prop:synch-converge} and Lemma~\ref{lemma:centralized-LS}.
%
\end{IEEEproof}

\subsection{PageRank computation in Google}

In this application, we study a network consisting of web pages~\cite{SB-LP:98}. This network can be represented by a graph $ {\G}=( {\V}, {\E})$, where the set of vertices correspond to the web pages and edges represent the links between the pages, i.e., the edge $(i,j)\in\mathcal{E}$, if page $i$ has an outgoing link to page $j$, or in other words, page $j$ has
an incoming link from page $i$. 

The goal of the PageRank algorithm is to provide a measure of relevance of each web page: the PageRank value of a page is a real number in $[0, 1]$, which is defined next.
Let us denote $\neigh_i=\setdef{h\in \V}{(i,h)\in \E}$ and $n_i=\card{\neigh_i}$, for each node $ i\in \V$, and $A$ the matrix such that 
\begin{align*}
A_{ij}=\begin{cases}
1/{n_j}&\text{if }j\in \neigh_i\\
0&\text{otherwise}. 
\end{cases}
\end{align*}
Let $m\in (0,1)$ and recall $n=\card{\V}$, and define 
\begin{equation}
M=(1-m) A+ \frac{m}n \1\1^{\top}.	
\end{equation} 
The PageRank of the graph $\G$ is the vector $\pgr$ such that $M\pgr=\pgr$ and $\sum_i\pgr_i=1$.

Given the initial condition such that $\1^{\top} x(0)=1$ (i.e., it is a stochastic vector), the PageRank vector can be computed through the recursion
\begin{equation}\label{eq:pow_meth}
x(k+1)=Mx(k)=(1-m)Ax(k)+\frac{m}{n}\1.
\end{equation}
In this case, we observe that the PageRank vector can be represented in terms of the affine dynamics (\ref{eq:synchro-algo})
simply taking 
\begin{equation}\label{Pu-def-pagerank}P=(1-m)A\quad\text{ and } \quad u=\frac{m}{n}\1.\end{equation}

Before showing the convergence of this recursion, which is studied in Proposition \ref{prop:convPR}, we present a simple technical lemma. 
Although the result has already been used in the literature (e.g. in~\cite{DA-GC-FF-AO:11}), we include a short proof for completeness. 
We recall that a matrix is said to be substochastic if it is nonnegative and the entries on each of its rows (or columns) sum up to no more than one. Moreover, every node corresponding to a row (or column) which sums to less than one is said to be a {\em deficiency} node. 

\begin{lemma}\label{lemma:substoch_stab}
Consider a  substochastic matrix $Q\in \real^{\V\times \V}$. If in the graph associated to $Q$ there is a path from every node to a deficiency node, then $Q$ is Schur stable.
\end{lemma}

\begin{IEEEproof}
First note that  $Q^k$ is substochastic for all $k$.  More precisely,  if we let $\mathcal{V}_k$ to be the set of deficiency nodes of $Q^k$,  then $\mathcal{V}_k\subseteq \mathcal{V}_{k+1}$ for every positive integer $k$. Moreover, there exists $k^\star$ such that 
$\mathcal{V}_{k^\star}=\V$, that is all nodes for $Q^{k^\star}$ are deficiency nodes.
Stability then follows by Gershgorin Circle Theorem.
\end{IEEEproof}

\begin{proposition}[Convergence of PageRank computation]\label{prop:convPR}
For any initial condition $x(0)\in\R^{\V}$ such that $\1^{\top} x(0)=1$, the sequence in \eqref{eq:pow_meth} converges to
$\pgr=(I-(1-m)A)^{-1}\frac{m}{n}\1.$
\end{proposition}

\begin{IEEEproof}
Since $A$ is column-stochastic and $m\in(0,1)$ then $P$ is a substochastic matrix and every node is a deficiency node. From Lemma~\ref{lemma:substoch_stab}, the matrix $P$ is Schur stable and from Proposition~\ref{prop:synch-converge} the dynamics in \eqref{eq:pow_meth} converges to $x^\star={(I-(1-m)A)^{-1}}\frac{m}{n}\1$.
Moreover, $\pgr=M\pgr=(1-m)A\pgr+\frac{m}{n}\1\1^{\top}\pgr=(1-m)A\pgr+\frac{m}{n}\1$, from which we conclude $x^{\star}=\pgr$.
\end{IEEEproof}

\subsection{Opinion dynamics in social networks}
In this application, we study a classical model introduced in~\cite{NEF-ECJ:99} to describe the effect of social influence and prejudices in the evolution of opinions in a population in the presence of the so-called stubborn agents. We briefly review and cast this model into the general framework of affine dynamics defined in \eqref{eq:synchro-algo}.

We consider a finite population $\V$ of interacting agents, whose {\it social network} of potential interactions is encoded by a graph $\G=(\V,\E)$, endowed with a self-loop  $(i,i)$ at every node. At time $k\in\integernonnegative$, each agent $i\in\V$ holds a {\it belief or opinion} about an underlying state of the world. We denote the vector of beliefs as $x(k)\in\real^\V$. An edge $(i,j)\in \E$ means that agent $j$ may directly influence the opinion of agent $i$.
 
Let $W\in \real^{\V\times\V}$ be a nonnegative matrix which defines the strength of the interactions ($W_{ij}=0$ if $(i,j)\not\in \E$) and $\Lambda$ be a diagonal matrix describing how sensitive each agent is to the opinions of the others based on interpersonal influences. We assume that $W$ is row-stochastic, i.e., $W\1=\1$ and we set $\Lambda=I-\diag(W)$, where $\diag(W)$ collects the self-weights given by the agents.
The dynamics of opinions $x(k)$ proposed in \cite{NEF-ECJ:99} is 
\begin{equation}
\label{eq:friedkin}
x(k+1)= \Lambda W x(k) + (I-\Lambda) v
\end{equation}
where $x(0)=v$ and $v\in \real^\V$. The vector $v$, which corresponds to the individuals' preconceived opinions, also appears as an input at every time step. This model falls under the class of 
affine dynamics (\ref{eq:synchro-algo}) simply taking 
\begin{equation}\label{Pu-def-opinions}P=\Lambda W\quad \text{and} \quad u=(I-\Lambda) v.\end{equation}
As a consequence of~\eqref{eq:friedkin} and Proposition \ref{prop:synch-converge}, the opinion profile at time $k$ is equal to 
$$
x(k)=\big ( (\Lambda W)^k +\sum_{h=0}^{k-1} (\Lambda W)^h (I-\Lambda) \big) v.
$$ 
The limit behavior of the opinions is described in the following 
convergence result.

\begin{proposition}[Convergence of opinion dynamics]
\label{prop:convergence-friedkin}
Assume that in the graph
associated to $W$ for any node $\ell\in \V$ there exists a path from
$\ell$ to a node $i$ such that  $W_{ii}>0$. Then, the opinions converge to
$$ \op:=(I-\Lambda W)^{-1}(I-\Lambda)v.$$
\end{proposition}

\begin{IEEEproof}
Due to the assumption, $\Lambda$ is a substochastic matrix. Then, $\Lambda W$ is substochastic also, and 
Schur stable by Lemma~\ref{lemma:substoch_stab}. Thus, the dynamics in~\eqref{eq:friedkin} converges to $ \op$. 
\end{IEEEproof}

We remark that our assumption on the existence of the path implies that each agent is influenced by at least one stubborn agent.
As shown in the proof, this is sufficient to guarantee the stability of the opinion dynamics.
In practice, it is reasonable to think that most agents in a social network will have some (positive) level of obstinacy $W_{ii}>0$.

%

\section{Ergodic randomized dynamics over networks}\label{sect:randomized}


In this section, we first present our main result about ergodicity properties of randomized dynamics over networks. Our result regards suitable randomized versions of the dynamics in \eqref{eq:synchro-algo}. 
Subsequently, we show applications of this result to localization of sensor networks, PageRank computation, and opinion dynamics in social networks, according to the problem statements made above.

We consider a sequence of independent identically distributed (i.i.d.) random variables $\{\theta(k)\}_{k\in\integernonnegative}$ taking values in a finite set $\Theta$.
Given a realization $\theta(k)$, $k\in\integernonnegative$, we associate to it a matrix $P(k)=P({\theta(k)})\in\real^{\V\times \V}$ and an input vector $u(k)=u(\theta(k))\in\real^\V$, obtaining a time-varying discrete-time dynamical system of the form
\begin{equation}
\label{eq:asynchro-algo}
 x(k+1)=P(k) x(k)+u(k)
\end{equation}
with initial condition $x(0) \in \real^\V $.
We observe that the state $\{x(k)\}_{k\in\integernonnegative}$ is a Markov process because, given the current position of the chain, the conditional distribution of the future values does not depend on the past values.

It may happen that the dynamics~\eqref{eq:asynchro-algo} oscillates persistently and fails to converge in a deterministic sense: this behavior is  apparent in the examples we are interested in.
In view of this fact, we want to give simple conditions which guarantee other types of probabilistic convergence. To this end, we provide some classical probabilistic convergence notions \cite{VB:95}.

The process $\{x(k)\}_{k\in\integernonnegative}$ is {\em ergodic} if there exists a random variable $x_{\infty} \in \real^\V $ such that almost surely
\begin{equation}\label{eq:ergodicity}
\lim_{k\to \infty}\frac{1}{k}\sum_{\ell=0}^{k-1}x({\ell})=\Exp[x_{\infty}].
\end{equation} 
The closely related definition of {\em{mean-square ergodicity}} instead requires  $$
\lim_{k\to \infty}\Exp\left[\left\|\frac{1}{k}\sum_{\ell=0}^{k-1}x({\ell})-\Exp[x_{\infty}]
\right\|_2^2\right]=0.
$$
The time-average in~\eqref{eq:ergodicity} is called Ces\'aro average or Polyak average in some contexts~\cite{BTP-ABJ:92}.
In what follows, we mostly focus on almost-sure ergodicity, although also mean-square ergodicity is mentioned. Indeed, it is often possible to deduce mean-square convergence from almost sure convergence: for instance, this implication is true for a uniformly bounded sequence of random variables, by the Dominated Convergence Theorem~\cite{VB:95}.
Our main analysis tool is the following result.

\begin{theorem}[Ergodicity of affine dynamics]
\label{thm:ergodic}
Consider the random process $\{x(k)\}_{k\in\integernonnegative}$ defined in~(\ref{eq:asynchro-algo}), where $\{P(k)\}_{k\in\integernonnegative}$ and $\{u(k)\}_{k\in\integernonnegative}$ are i.i.d.\ and have finite first moments. If there exists $\alpha\in(0,1]$ such that 
\begin{equation}\label{eq:hyp}\Exp[P(k)]=(1-\alpha)I+\alpha P,\qquad\Exp[u(k)]=\alpha u,\end{equation} where $P$ and $u$ are given in Proposition \ref{prop:synch-converge}, then
\begin{enumerate}
\item $x(k)$ converges in distribution to a random variable $x_{\infty}$, and the distribution of $x_{\infty}$ is the unique invariant distribution for \eqref{eq:asynchro-algo};
\item the process is {\em ergodic};
\item the limit random variable satisfies
$
\Exp[x_\infty]=x^{\star}.
$ 
\end{enumerate}
\end{theorem}
The proof of Theorem~\ref{thm:ergodic} is technical and is postponed to the Appendix. We instead make the following first-order analysis of~\eqref{eq:asynchro-algo} under the assumptions of the Theorem.
Since $P$ is Schur stable in Proposition~\ref{prop:synch-converge}, so is $\Exp[P(k)]$ under the hypothesis~\eqref{eq:hyp}, and the expected dynamics of the process~\eqref{eq:asynchro-algo} can be interpreted as a ``lazy'' (slowed down) version of the synchronous dynamics~\eqref{eq:synchro-algo} associated to the matrix $P$. Indeed, 
 \begin{align*}
\Exp[x(k+1)]&=\Exp[P(k)]\Exp[x(k)]+\alpha\Exp[u(k)]\\
&=((1-\alpha)I+\alpha P)\Exp[x(k)]+\alpha u,
\end{align*}
from which
$\lim_{k\rightarrow+\infty}\Exp[x(k)]=x^{\star}$. 

The following refinement of Theorem~\ref{thm:ergodic} --proved in Appendix~\ref{sec:proofB}-- is specially useful to our purposes.
\begin{corollary}[Ergodicity of affine dynamics on random subsequences]
\label{cor:ergodic2}
Consider the random process $\{x(k)\}_{k\in\integernonnegative}$ defined in~(\ref{eq:asynchro-algo}), where $\{P(k)\}_{k\in\integernonnegative}$ and $\{u(k)\}_{k\in\integernonnegative}$ are i.i.d.\ and have finite first moments. 
Let $\{\omega(k)\}_{k\in\integernonnegative}\in\{0,1\}^{\integernonnegative}$ be an i.i.d. random sequence such that, for all $k$, $\omega(k)$ is independent of $P(\ell)$ for all $\ell<k$ and  $\omega(k)\neq 0$ with positive probability.
If there exists $\alpha\in(0,1]$ such that 
\begin{equation}\label{eq:hyp}\Exp[P(k)]=(1-\alpha)I+\alpha P,\qquad\Exp[u(k)]=\alpha u,\end{equation} where $P$ and $u$ are given in Proposition~\ref{prop:synch-converge}, then almost surely
$$
\lim_{k\rightarrow+\infty}\frac{1}{\sum_{i=0}^{k-1}\omega(i)}\sum_{\ell=0}^{k-1}\omega(\ell)x(\ell)=x^{\star}.
$$
\end{corollary}

In the rest of this section, Theorem~\ref{thm:ergodic} and Corollary~\ref{cor:ergodic2} will be applied to specific dynamics in sensor localization, PageRank computation, and opinion dynamics.

\subsection{Sensor localization in wireless networks (cont'd)}\label{sect:random-localization}

This section is devoted to describe a randomized algorithm, which was proposed in~\cite{CR-PF-HI-RT:13a} to solve the sensor localization problem.
For each node $i\in \V$, the algorithm involves a triple of states $(x_i,\kappa_i,\tilde x_i)$, which depend on a discrete time index $k\in \integernonnegative$. These three variables play the following roles: $x_i(k)$ is the ``raw'' estimate of $s_i$ obtained by $i$ at time $k$ through communications with its neighbors, $\kappa_i(k)$ counts the number of updates performed by $i$ up to time $k$, and $\tilde x_i(k)$ is the ``smoothed'' estimate obtained through time-averaging.
The algorithm is defined by choosing a scalar parameter $\gamma\in(0,1)$ and a sequence of random variables $\{\theta(k)\}_{k\in \integernonnegative}$ taking values in $\E$. 
The state variables are initialized as $(x_i(0),\kappa_i(0),\tilde x_i(0))=(0,0,0)$ for all $i$, and at each 
time $k>0$, provided that $\theta(k)=(i,j)$, 
the states are updated according to the following recursions, namely the raw estimates as
\begin{subequations} \label{dyn2}
\begin{equation} \label{dyn2a}
\begin{split}
x_i(k+1)&=(1-\gamma)x_i(k)+ \gamma x_j(k)+ \gamma b_{(i,j)}\\
x_j(k+1)&=(1-\gamma)x_j(k)+ \gamma x_i(k)-\gamma b_{(i,j)}\\
x_{\ell}(k+1)&=x_{\ell}(k)\qquad \text{if }\ell\notin\{i,j\};\\
\end{split}
\end{equation}
the local times as
\begin{equation}
\begin{split}
\kappa_i(k+1)&=\kappa_i(k)+1\\
\kappa_j(k+1)&=\kappa_j(k)+1\\
\kappa_{\ell}(k+1)&=\kappa_{\ell}(k)\qquad \text{if }\ell\notin\{i,j\};\\
\end{split}
\end{equation}
and the time-averages as
\begin{equation}
\begin{split}\label{dyn2b}
\widetilde x_i(k+1)&=\frac1{\kappa_i(k+1)}\big( \kappa_i(k) \widetilde x_i(k) + x_i(k+1)\big)\\
\widetilde x_j(k+1)&=\frac1{\kappa_j(k+1)}\big( \kappa_j(k) \widetilde x_j(k) + x_j(k+1)\big)\\
\widetilde x_{\ell}(k+1)&=\widetilde x_{\ell}(k)\qquad \text{if }\ell\notin\{i,j\}.
\end{split}
\end{equation}
\label{eq:gossip-algo-ki}
\end{subequations}
Next, we assume the sequence $\{\theta(k)\}_{k\in \integernonnegative}$ to be i.i.d., and its probability distribution to be {\it uniform}, i.e.,
\begin{equation}\label{eq:uniform}
\Prob[\theta(k)=(i,j)]=\frac{1}{|\mathcal{E}|}, \qquad\forall k\in\integernonnegative.
\end{equation}
Note that this choice is made for simplicity, but this approach may easily accommodate other distributions if required by specific applications.

\begin{remark}[Local and global clocks]
It should be noted that the time index $k$ counts the number of updates which have occurred in the network, whereas for each $i\in \V$ the variable $\kappa_i(k)$ is the number of updates involving $i$ up to the current time. Hence, $\kappa_i$ is a local variable which is inherently known to agent $i$, even in case a common clock $k$ is unavailable.
Therefore, this algorithm is {\it totally asynchronous} and {\it fully distributed}, in the sense that the updates, including the time-averaging process, do not require the nodes to be aware of a common clock. This feature is especially attractive if the algorithm has to be applied to clock synchronization problems.
\end{remark}

\begin{figure}[h]
\begin{center}
\psfrag{x2}[][][1][-90]{$x$}
\psfrag{x4}[][][1][-90]{$\widetilde x$}
 \psfrag{k}{$k$}
\includegraphics[width=0.85\columnwidth]{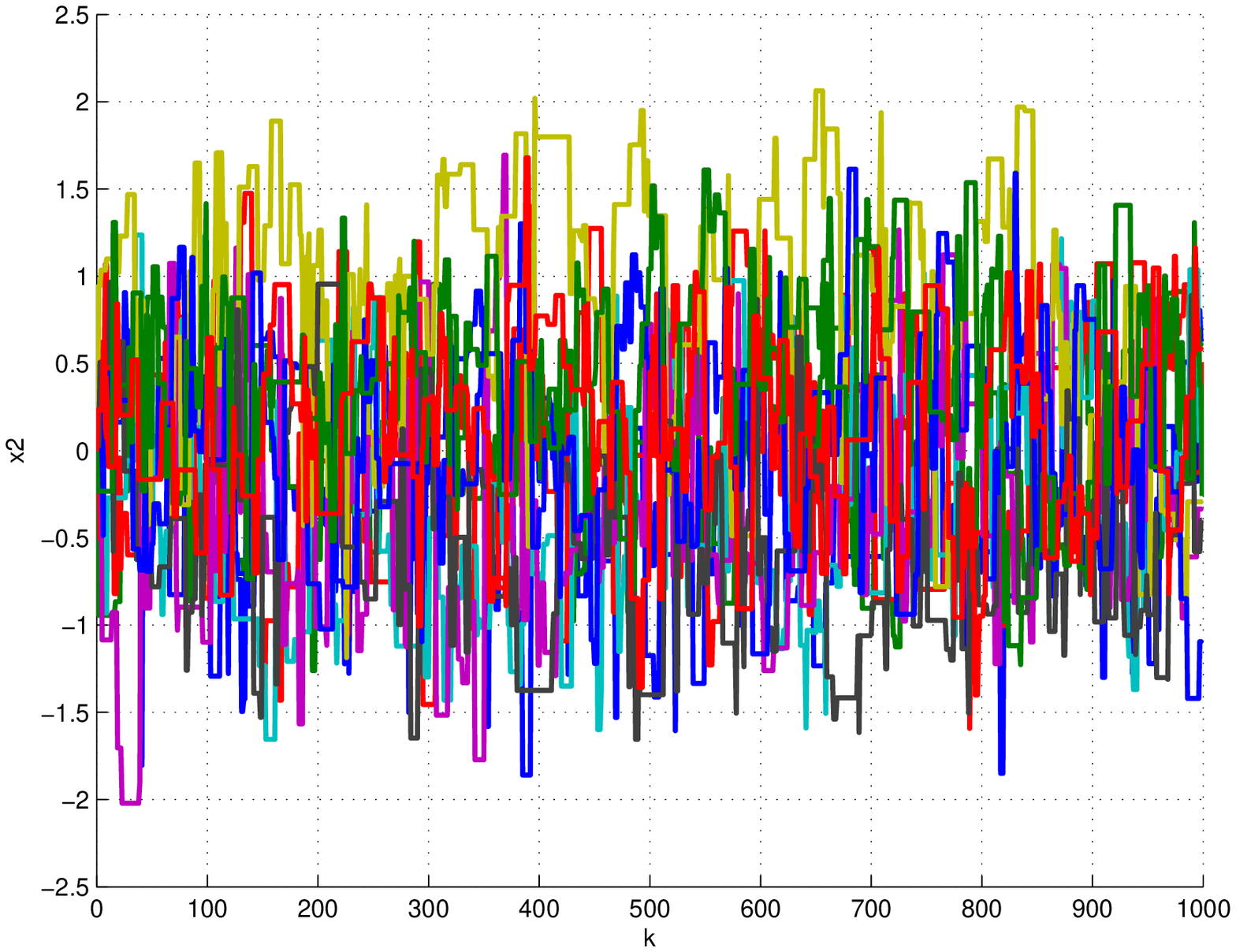}
\includegraphics[width=0.85\columnwidth]{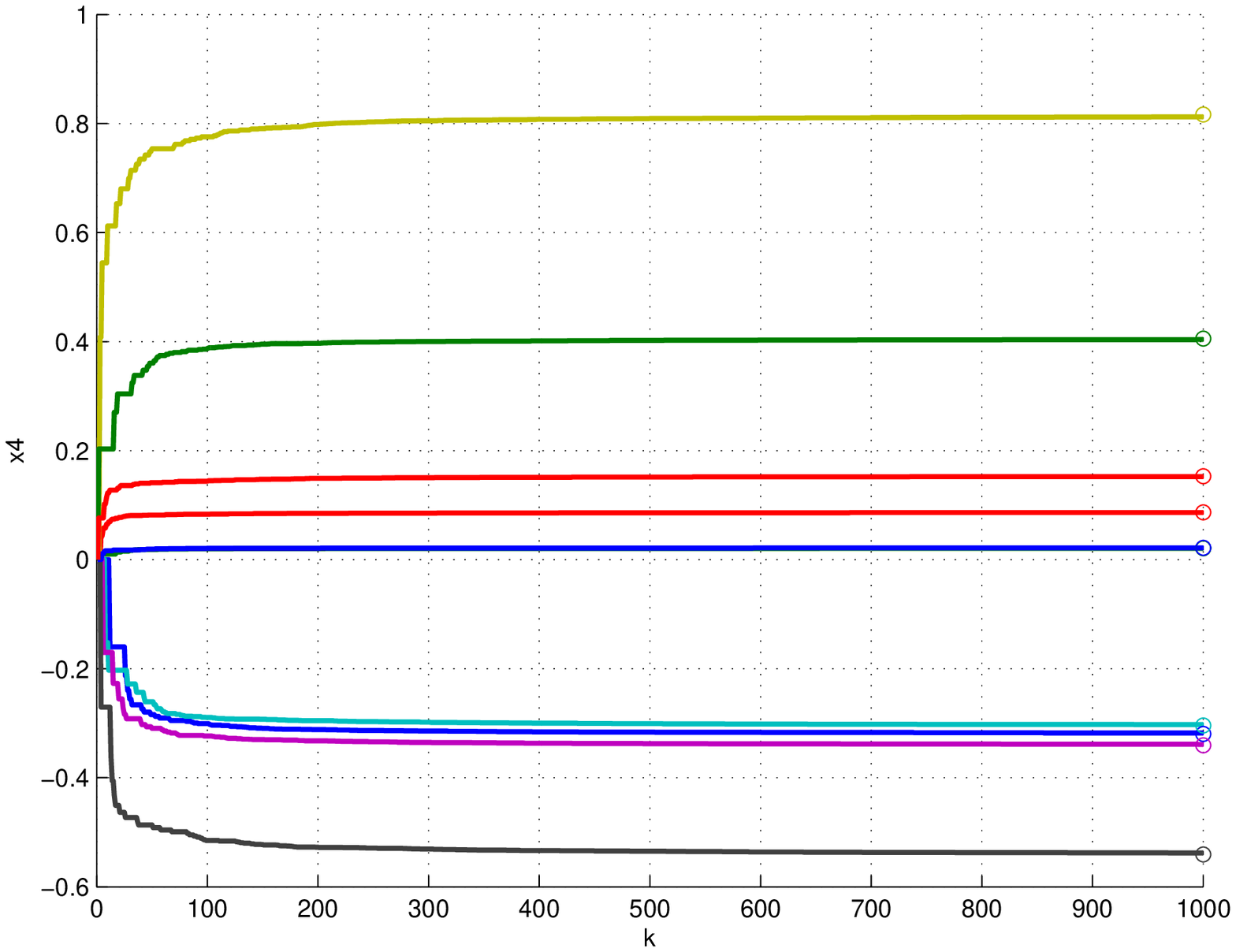}
\caption{Dynamics~\eqref{dyn2} running on a complete graph with $10$ nodes, with $\gamma=0.5$.}
\label{fig:simul-complete}
\end{center}
\end{figure}

The dynamics in \eqref{dyn2a} oscillates persistently and fails to converge in a deterministic sense, as shown in Figure~\ref{fig:simul-complete} for a complete graph\footnote{A 
complete graph is
an graph in which every pair of distinct vertices is connected by an edge.}.
However, the oscillations asymptotically concentrate around the solution of the least-squares problem, as it is formally stated in the following result, which shows that the sample dynamics is well-represented by the average one. This 
indicates that $\widetilde x_i(k)$ is ``the right variable'' to approximate the optimal estimate $\loc$ because the process $x(k)$ is ergodic. In the proof of the theorem, we show that the dynamics in equation (\ref{dyn2a})
can be written in terms of the more general process (\ref{eq:asynchro-algo}). 

\begin{theorem}[Ergodicity of sensor localization]
\label{theorem:convforlocal}
The dynamics in~\eqref{dyn2} with uniform selection \eqref{eq:uniform} is such that $\lim_{k\to\infty} \widetilde{x}(k)=\loc$
almost surely. 
\end{theorem}
\begin{IEEEproof}
We rewrite the dynamics of~\eqref{dyn2a} as
\begin{equation}\label{eq:dyn3-old}
x(k+1)=Q(k)x(k)+u(k)
\end{equation}
and, provided $\theta(k)=(i,j)$, we define
$$
Q(k)=I-\gamma(e_i-e_j)(e_i-e_j)^\top
$$ 
and
$
u(k)=b_{\theta(k)}(e_i-e_j),
$ 
where the vector $e_i$ is defined in the preliminaries.
We note that  for all $k$ the matrix $Q(k)$ is doubly stochastic and the sum of the elements in $u(k)$ is zero:
in particular, given $x(0)=0$, then $\1^\top x(k)=0$ for each $k\in\integernonnegative$.
These observations further imply that the dynamics of $x(k)$ is equivalently described by the iteration
\begin{equation}
\label{eq:dyn3}
x(k+1)=Q(k)\Omega x(k)+u(k),
\end{equation}
where $\Omega=I-\frac1n\1\1^\top$ as previously in Subsection~\ref{es1}.
%
Letting $P(k)=Q(k)\Omega$, the dynamics of the algorithm is cast in the form of \eqref{eq:asynchro-algo}.
Next, using the uniform distribution~\eqref{eq:uniform}, we compute
\begin{gather*}
\Exp[P(k)]=\left(I-\gamma\frac{L}{|\mathcal{E}|}\right)\Omega,
\\
\Exp\left[ u(k)\right] = \gamma\frac{A^\top b}{|\mathcal{E}|},
\end{gather*}
and observe that $\Exp[P(k)]$ satisfies ergodicity condition in Theorem~\ref{thm:ergodic} with $P$ and $u$ defined in~\eqref{Pu-def-localization}, $\alpha=1$ and $\tau=\gamma/\card{\E}$. 
If we define, for all $i\in\V$ and all $k\in\integernonnegative$, $$\omega_i(k)=\begin{cases}
1&\text{if }\theta(k)=(i,j) \text{ or }\theta(k)=(j,i)\\
0&\text{otherwise}
\end{cases}$$
then
$\kappa_{i}(k+1)=\kappa_i(k)+\omega_i(k)=\sum_{\ell=0}^{k}\omega_i(\ell)$
and
$$
\widetilde{x}_{i}(k+1)=\frac{1}{\sum_{\ell=0}^{k}\omega_i(\ell)}\sum_{\ell=0}^{k}\omega_i(\ell)x_i(\ell).
$$
Being $\{\omega(k)\}_k$ an i.i.d. random sequence, by Corollary~\ref{cor:ergodic2} we conclude our argument.
\end{IEEEproof}

\begin{remark}[Noise-free measurements]\label{rem:noise-free-localization}
It is easy to see that if measurements have no noise ($\sigma^2=0$), then $x(k)$ itself converges to the exact solution $\loc$, and moreover convergence is exponentially fast. This fact is also proved in~\cite{NMF-AZ:12}.
\end{remark}

\begin{remark}[Mean-square ergodicity]
It is also true that $\widetilde{x}(k)$ converges to $\loc$ in the mean-square sense. 
A proof can be obtained with similar arguments as in~\cite{CR-PF-HI-RT:13a} and is not detailed here.
\end{remark}

\subsection{PageRank computation in Google (cont'd)}\label{sect:random-pagerank}

We now describe a new example of an ``edge-based'' randomized gossip algorithm. Its motivation comes from the interest in reducing the communication effort required by the network. Being only one edge activated at each time, such effort is minimal. 

Each node $i\in\V$ holds a couple of states $(x_i,\overline{x}_i)$. For every time step $k$ an edge $\theta(k)$ is sampled from a uniform distribution over $\E$ (note that sampling is independent at each time $k$).
Then, the states are updated as follows:
\begin{subequations}\label{eq:gossip-pr1}
\begin{align} 
x_i(k+1)=&(1-r) \left(1-\frac1{n_i}\right)\, x_i(k)+ \frac{r}{n}\\
x_j(k+1)=&(1-r) \left(x_j(k)+\frac1{n_i} x_i(k)\right)+ \frac{r}{n}\\
x_h(k+1)=&(1-r) \, x_h(k)+ \frac{r}{n} \qquad \text{if}\; h\neq i,j \label{eq:gossip-pr1-3}
\end{align}
\end{subequations}
and
\begin{align} \label{eq:gossip-pr2}
\overline{x}_\ell(k+1)=\frac{kx_\ell(k)+x_\ell(k+1)}{k+1} \quad \forall\,\ell\in \V
\end{align}
where $r\in(0,1)$ is a design parameter to be determined.
The update in \eqref{eq:gossip-pr1} can also be formally rewritten in vector-wise form as
$$ 
x(k+1)=P(k) x(k)+u(k),
$$
where 
$$
P(k)=(1-r) A(k),\quad u(k)= \frac{r}n \1.
$$ 
Here $A(k)$ and $P(k)$ are random matrices which are determined by the choice of $\theta(k)=(i,j)$
$$ 
A(k)=I+\frac1{n_i}(e_je_i^\top-e_ie_i^\top).
$$
Then, $A(k)$ is uniformly distributed over the set of matrices $\setdef{I+\frac1{n_i}(e_je_i^\top-e_ie_i^\top)}{(i,j)\in \E}$.

\begin{remark}[Local and global clocks]
We note that, opposed to~\eqref{dyn2}, algorithm~\eqref{eq:gossip-pr1} does require the nodes to access the global time variable $k$. The reason for this synchrony requirement comes from the need to preserve the stochasticity of the vector $x(k)$, which is guaranteed by~\eqref{eq:gossip-pr1-3}. We believe this is a reasonable assumption, because these algorithms are to be implemented on webpages or domain servers which are typically endowed with clocks.
\end{remark}

In the next result, we state convergence of this algorithm.
\begin{theorem}[Ergodic PageRank convergence]
\label{cor:pagerank}
Let us consider the dynamics~\eqref{eq:gossip-pr1}-\eqref{eq:gossip-pr2} with 
$$
r=\frac{m}{m-\card{\E}m+\card{\E}}
$$ 
where $x(0)$ is a stochastic vector. Then, the
sequence $\{\overline{x}(k)\}_{\integernonnegative}$ 
is such that 
$
\lim_{k\to\infty} \overline{x}(k)=\pgr
$
almost surely. 
\end{theorem}

\begin{IEEEproof}
For each $k\in\integernonnegative$, we have
\begin{align*}
\Exp[A(k)]&=\frac{1}{\card{\E}}\sum_{(i,j)\in \E}\left(I+\frac1{n_i}(e_je_i^\top-e_ie_i^\top)\right)\\
&=I+\frac{1}{\card{\E}}\sum_{(i,j)\in \E} \frac1{n_i} e_je_i^\top-\frac{1}{\card{\E}}\sum_{(i,j)\in \E} \frac1{n_i} e_ie_i^\top\\
&=I+\frac{1}{\card{\E}}\sum_{(i,j)\in \E} \frac1{n_i} e_je_i^\top-\frac{1}{\card{\E}}\sum_{i\in V} \sum_{j\in \neigh_i} \frac1{n_i} e_ie_i^\top\\
&=I+\frac{1}{\card{\E}}\sum_{(i,j)\in \E} A_{ji}e_je_i^\top-\frac{1}{\card{\E}}\sum_{i\in V} \frac{n_i}{n_i} e_ie_i^\top\\
&=\left(1-\frac1{|\E|}\right) I+ \frac1{|\E|} A.
\end{align*}
It should be noted that, setting $\alpha=({m-m\card{\E}+\card{\E}})^{-1}$ and $P$ and $u$ as in~\eqref{Pu-def-pagerank},
\begin{align*}
\Exp[P(k)]&=(1-r)\Exp[A(k)]\\
&=\left(1-\alpha\right)I+\alpha (1-m)A\\
&=\left(1-\alpha\right)I+\alpha P,
\end{align*}
and $\Exp[u(k)]=\alpha\frac{m}{n}\1=\alpha u.$
From Theorem~\ref{thm:ergodic} we conclude convergence almost surely. 
\end{IEEEproof}

This result can also be proved by techniques from stochastic approximation. Such techniques have already been effectively applied to specific algorithms for PageRank computation~\cite{DBLP:journals/corr/abs-1305-3178}. 

\begin{remark}[Mean-square ergodicity]\label{rem:ms-pgr}
Since $x(k)$ are stochastic vectors, they are uniformly bounded and by the Dominated Convergence Theorem we conclude the convergence in the mean-square sense. Mean-square ergodicity of randomized PageRank was already proved in~\cite{HI-RT-EWB-FD:09} under assumptions which are equivalent to those in Theorem~\ref{thm:ergodic}.
\end{remark}

\subsection{Opinion dynamics in social networks (cont'd)}
\label{sect:random-friedkin}
In this subsection, we introduce a randomized model of the communication process among the agents in the Friedkin and Johnsen's model presented above. As a result, we obtain a new class of randomized opinion dynamics.

The problem is now described.
Each agent $i\in \V $ possesses an initial belief $x_i(0)=v_i\in \real$, as in the model~\eqref{eq:friedkin}. At each time $k\in\integernonnegative$ a link is randomly sampled from a uniform distribution over $\E$. If the edge $(i, j)$ is selected at time $k$, agent $i$ meets agent $j$ and updates its belief to a convex combination of its previous belief, the belief of $j$, and its initial belief. Namely, 
\begin{align}
\label{eq:gossip-friedkin}
\nonumber x_i(k+1)&=h_{i}\big((1-\Gamma_{ij})x_i(k)+\Gamma_{ij}x_j(k)\big)+(1-h_{i})v_i\\
x_\ell(k+1)&=x_\ell(k)\qquad \forall \ell\in \V\setminus\{i\},
\end{align}
where the weighting coefficients $h_i$ and $\Gamma_{ij}$ are defined as 
\begin{equation} 
\label{eqn:H}
 h_i=\begin{cases}
1-{(1-\lambda_{i})}/{d_i}&\text{if }d_i\neq1\\
{ 0}&\text{otherwise}
\end{cases}
\end{equation}
\begin{equation}\label{eqn:Gamma}
\Gamma_{ij}=\begin{cases}
\frac{d_i(1-h_i)+h_i-(1-\lambda_{i}W_{ii})}{h_i}&\text{if }i=j,\ d_i\neq1\\
\frac{\lambda_{i}W_{ij}}{h_i}&\text{if }i\neq j,\ d_i\neq 1\\
1&\text{if }i=j,\ d_i=1\\
0&\text{if }i\neq j,\ d_i=1
\end{cases}
\end{equation}
where the matrices $W$ and $\Lambda$ are those in~\eqref{eq:friedkin} and $d_i=|\setdef{h}{(i,h)\in \E}$. Recall that $d_i\ge1$ by the presence of self-loops. 
It is immediate to observe that
(a) $h_i\in [0,1]$ for all $i\in \V$; (b) $\Gamma$ is adapted to the graph $\mathcal{G}$;
(c) $\Gamma$ is row-stochastic; and (d) at all times the opinions of the agents are convex combinations of their initial prejudices.

We now study the convergence properties of the gossip opinion dynamics and we show that the opinions converge to the same value $\op$ given in Proposition~\ref{prop:convergence-friedkin}. 
In the proof of the result, we show that the dynamics in equation~\eqref{eq:gossip-friedkin}
can be written in terms of the more general process (\ref{eq:asynchro-algo}).

\begin{theorem}[Ergodic opinion dynamics]
\label{thm:gossip-opinions} 
Assume that in the graph
associated to W for any node $\ell\in \V$ there exists a path from
$\ell$ to a node $i$ such that  $W_{ii}>0$.
Then, the dynamics~\eqref{eq:gossip-friedkin} is 
almost surely and mean-square ergodic, and the time-averaged opinions defined in \eqref{eq:ergodicity} converge to $\op$.
\end{theorem}
\begin{IEEEproof}
Provided the edge $\theta(k)=(i,j)$ is chosen at time $k$, the dynamics~\eqref{eq:gossip-friedkin} can be rewritten in vector form as
\begin{align*}
x(k+1)=&(I-e_ie_i^\top(I-H))\left(I+\Gamma_{ij}(e_ie_j^\top-e_ie_i^\top)\right)x(k)\\&+e_ie_i^\top(I-H)v.
\end{align*}
If we define the matrices 
\begin{align*}P^{ij}=& (I-e_ie_i^\top(I-H))\left(I+\Gamma_{ij}(e_ie_j^\top-e_ie_i^\top)\right)\\
u^{ij}=&\, e_ie_i^\top(I-H)v,\end{align*}
then the dynamics is 
$
x(k+1)=P^{ij} x(k) + u^{ij}.
$
Note that
the expressions in \eqref{eqn:H} and \eqref{eqn:Gamma} imply
\begin{gather*}
D(I-H)=I-\Lambda\\
D(I-H)+H(I-\Gamma)=I-\Lambda W
\end{gather*}
where $H = {\rm diag}\{h_1, h_2, \ldots, h_n\}$ and $D= {\rm diag}\{d_1, d_2, \ldots, d_n\}$. Consequently, one can compute the generic entries of the expected matrix $\Exp[P(k)]=\frac1{\card{\E}} \sum_{(\ell,m)\in\E} P^{\ell m}$ as 
$\Exp[P(k)]_{ij}=\frac{1}{\card{\E}} h_i \Gamma_{ij}=\frac{1}{\card{\E}}\lambda_iW_{ij}$ if $i\neq j$,   and
\begin{align*}\Exp[P(k)]_{ii}&=1-\frac{1}{\card{\E}}\big(   d_i(1-h_i)+h_i(1-\Gamma_{ii})\big)\\
&=\Big(1-\frac{1}{\card{\E}}\Big)+ \frac{1}{\card{\E}} \lambda_iW_{ii} .\end{align*} From these formulas, we conclude that $\Exp[P(k)]=(1-\frac{1}{\card{\E}})I+ \frac{1}{\card{\E}} \Lambda W$ and $\Exp[u(k)] = \frac{1}{\card{\E}} (I-\Lambda)v$.
Then, using~\eqref{Pu-def-opinions} the claim follows by Theorem~\ref{thm:ergodic} and Proposition~\ref{prop:convergence-friedkin}.
\end{IEEEproof}

\begin{figure}[t]
\begin{center}
\psfrag{x}[][][1][-90]{$x$}
\psfrag{averaged}[][][1][-90]{$\bar x$}
 \psfrag{time}{$k$}
\includegraphics[width=0.85\columnwidth]{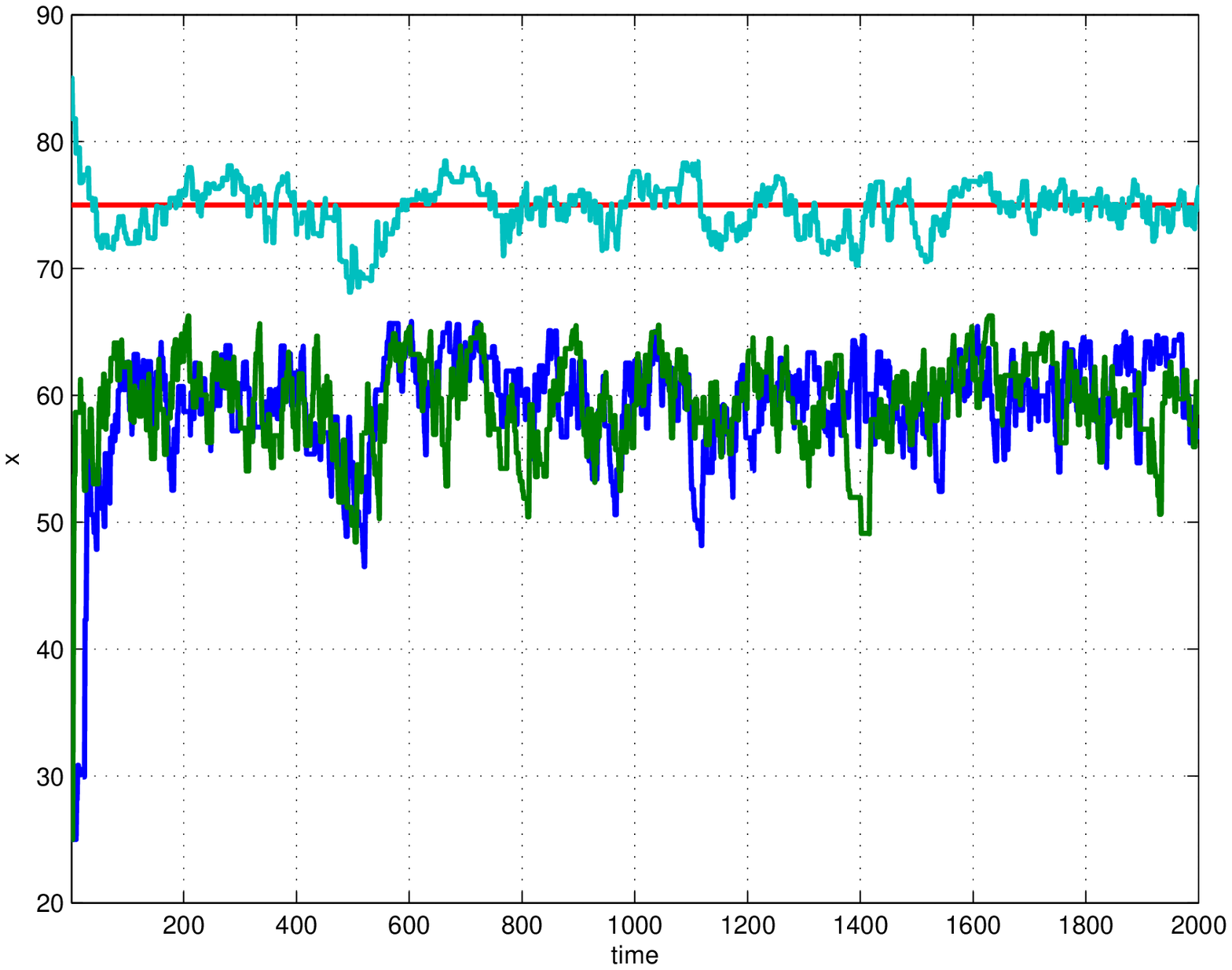}
\includegraphics[width=0.85\columnwidth]{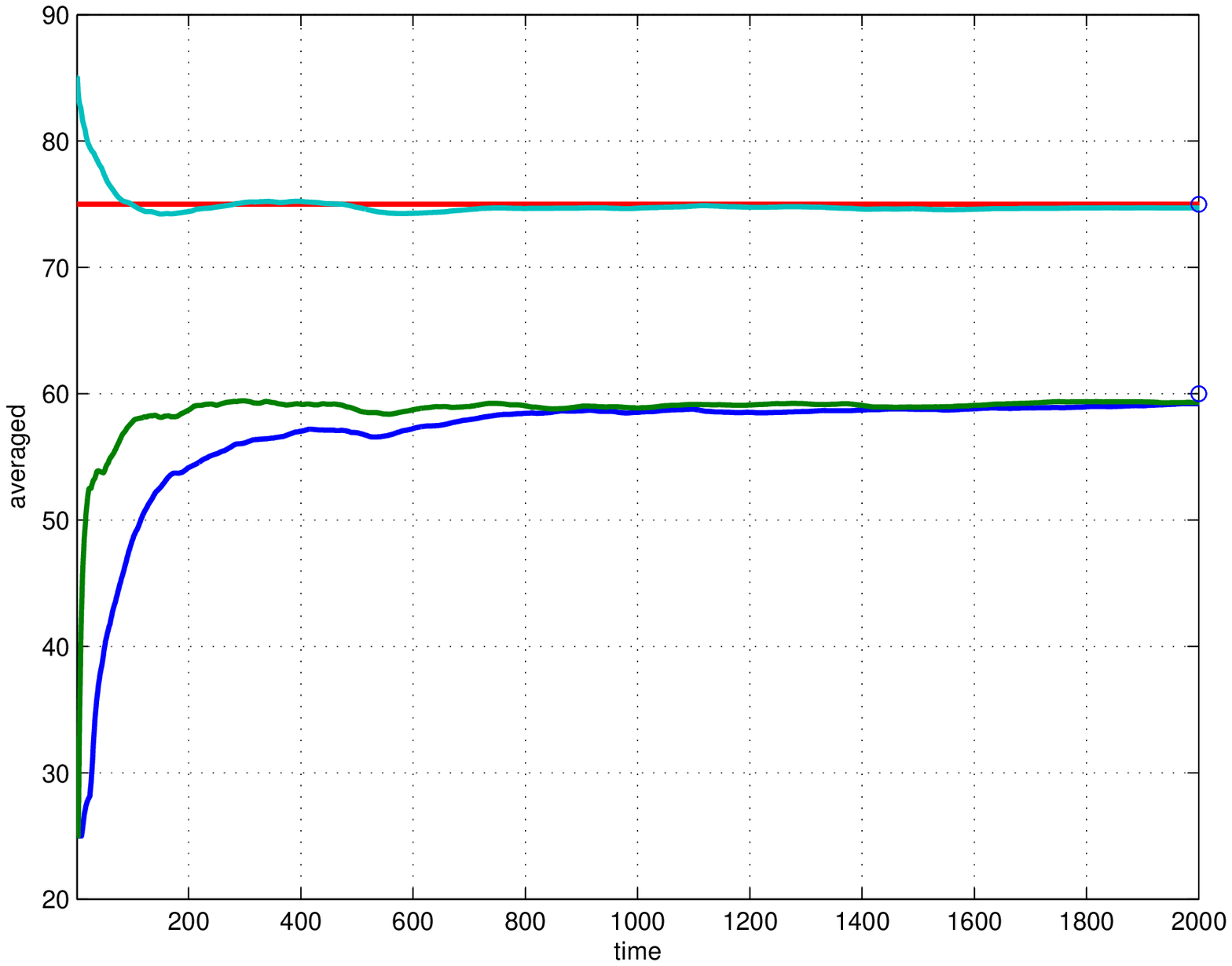}
\caption{Four-nodes social network from~\cite{NEF-ECJ:99}. The opinion process
$x$ (top plot) oscillates persistently. As the belief process is ergodic, the time-averages $\bar x$ (bottom plot) converge, when time goes to
infinity, to $\op$ (marked by blue circles).}
\label{fig:simul-od}
\end{center}
\end{figure}

\begin{remark}[Mean-square ergodicity]
Since the opinions are uniformly bounded, by the Dominated Convergence Theorem we  also conclude the convergence in the 
mean-square sense. Mean-square ergodicity is also proved in~\cite{PF-CR-RT-HI:13c} under assumptions which are equivalent to those in Theorem~\ref{thm:ergodic}.
\end{remark}
The ergodicity of the opinion dynamics is illustrated by the simulations in
Figure \ref{fig:simul-od}, which plots the state $x(k$) and the
corresponding time-averages, respectively.

We notice that the dynamics~\eqref{eq:gossip-friedkin} assume the edges to be chosen for the update according to a uniform distribution. This choice is made for simplicity, but our analysis can easily be extended to consider more general or different distributions.

We now discuss the interpretation of our convergence theorem in the
context of opinion dynamics. The original model by Friedkin and Johnsen abstracts from a precise analysis of the communication process among the agents, and postulates synchronous rounds of interaction. In fact, the lack of a more precise model for inter-agent interactions is acknowledged in~\cite{NEF-ECJ:99} by saying that ``it is obvious that interpersonal influences do not occur in the simultaneous way that is assumed''.
Our gossip dynamics introduces a more realistic model of  the communication process among the agents: indeed the agents were allowed to discuss asynchronously and in pairs in the experiments reported in~\cite{NEF-ECJ:99}. 
We believe that the relationship between the randomized and the synchronous dynamics provides an additional justification and a new perspective on the model originally proposed by Friedkin and Johnsen: an example of social network derived from their experiments is analysed in detail in~\cite{PF-CR-RT-HI:13c}.

The forms of~\eqref{eqn:H} and \eqref{eqn:Gamma} may seem complicated at first sight. However, this is not surprising if we think of other examples of randomized dynamics over networks. For instance, in problems of consensus~\cite{FF-SZ:08a}, localization~(see Section \ref{sect:random-localization}), and PageRank computation~(see Section \ref{sect:random-pagerank}), the definition of the update matrices of the randomized dynamics is not trivial and must be done carefully in order to reconstruct, on average, the desired synchronous dynamics.
 From a sociological perspective, \eqref{eqn:H} and~\eqref{eqn:Gamma} postulate a specific form of interaction for individuals in pairwise meetings, which is reflected on average by Friedkin and Johnsen's dynamics. Since by~\eqref{eqn:H} $h_i>\lambda_{ii}$, we observe that individuals display a lower obstinacy during pairwise interaction.

\section{Concluding remarks}\label{sect:conclusion}
In this work, we have proposed time-averaging as a tool for smoothing oscillations in randomized network systems.  Other authors have proposed different solutions, which damp the inputs to the dynamics in the long run: this goal is achieved through ``under-relaxations'', that is, by using gains (or equivalently step-sizes) which decrease with time.
The analysis of the resulting dynamics is often based on tools from stochastic approximation~\cite{VSB:08} or semi-martingale theory~\cite[Ch~2]{BP:87}. Notably, also the choice of decreasing gains can be performed asynchronously and without coordination, albeit at the price of a more complex analysis~\cite[Ch.~7]{VSB:08} \cite{AN:11}.

Our method of time-averaging, together with its analysis based on ergodicity, has three advantages:  (i) it is simple to apply as it requires minimal assumptions, (ii) it allows for a unified treatment of different algorithms, and (iii) it gives a qualitative insight into the stochastic processes of interest. 
However, the use of time-averaging is not itself free from drawbacks. Indeed, rate of convergence of time-averages is not exponential, as for the original synchronous dynamics, but polynomial ($O(1/k)$).
This fact can be observed by inspecting the proof of Theorem~\ref{thm:ergodic} or from a mean-square convergence analysis, as we did in~\cite{HI-RT:10} and~\cite{CR-PF-HI-RT:13a}.
This drawback, which is shared by over-relaxation approaches, stimulates research towards exponentially-fast algorithms. Likely, effective algorithms can be constructed by endowing the nodes with some memory capabilities: an example is provided in~\cite{RC-AC-LS-MT:13} for the localization problem. More generally, their design may be based on the so-called asynchronous iteration method from numerical analysis~\cite[Section 6.2]{DPB-JNT:89}: for instance, the application of this method to PageRank computation is discussed in~\cite[Section~VII]{HI-RT:10}.

Finally, we expect that the approach presented here can be applied to a wide range of problems in network systems, besides the three examples detailed here. Venues for application include gossip algorithms to solve problems of simultaneous estimation and classification in sensor networks~\cite{FF-SF-CR:11}, convex optimization~\cite{AN:11}, and optimization in power networks~\cite{SB-SZ:13}.

\section*{Acknowledgements}
The authors are grateful to Noah E. Friedkin for posing the problem studied in Section~\ref{sect:random-friedkin} and to Keyou You for finding an error in a preliminary version of this paper~\cite{CR-PF-RT-HI:13b} and for providing a copy of~\cite{KY-SS-LQ:14}. The authors would also like to thank Francesco Bullo, Giacomo Como, and Fabio Fagnani for interesting conversations on the topics of this paper.

\appendices
\section{Proof of Theorem~\ref{thm:ergodic}}
\label{sec:proof}
In this appendix, we provide the proof of our main result regarding randomized dynamics defined in~\eqref{eq:asynchro-algo}. 
%
The proof is based on techniques for iterated random functions, which we recall from~\cite{PD-DF:99}. These techniques require, in order to study the random process~\eqref{eq:asynchro-algo}, to consider the associated {\em backward} process $\overleftarrow{x}(k)$, which we define below.

For any time instant $k$, consider the random matrices $P{(k)}$ and ${u}{(k)}$ and define the matrix product
\begin{equation}\label{P}
\overrightarrow{P}(\ell,m):=P{(m)}P{(m-1)}\cdots P{(\ell+1)}P{(\ell)}
\end{equation}
with $\ell\in\{0,\ldots,m\}.$
Then, the iterated affine system in~\eqref{eq:dyn3} can be rewritten as
\begin{equation}
 x(k+1)=\overrightarrow{P}(0,k)  x(0)+\sum_{0\leq \ell\leq k}\overrightarrow{P}(\ell+1,k)u(\ell).
\end{equation}
The corresponding \emph{backward process} is defined by
\begin{equation}
\overleftarrow{x}(k+1)=\overleftarrow{P}(0,k)  x(0)+\sum_{0\leq \ell\leq k}\overleftarrow{P}(0,\ell-1)u(\ell),
\end{equation}
where
\begin{equation}\label{Preverse}
\overleftarrow{P}(\ell,m):=P{(\ell)}P{(\ell+1)} \cdots P{(m-1)} P{(m)} 
\end{equation}
with $\ell\in\{0,\ldots,m\}.$
Crucially, the backward process $\overleftarrow{x}(k)$ has at every time $k\in\integernonnegative$ the same probability distribution as $x(k)$. 
The main tool to study the backward process is the following result. Let $\|\cdot\|$ denote any norm.

\begin{lemma}[Theorem~2.1 in~\cite{PD-DF:99}]\label{Diaco}
Let us consider the Markov { {process}} $\{x(k)\}_{k\in\integernonnegative}$ defined by
$$x({k+1})={P(k)}x(k) + u(k)\quad k\in \integernonnegative$$ 
where $P(k)\in\R^{\V\times\V}$ and $u(k)\in\R^{\V}$ are i.i.d. random variables.
Let us assume that
\begin{equation}
\Exp[\log\|P(k)\|]<\infty\qquad
\Exp[\log\|u(k)\|]<\infty.
\end{equation}
The corresponding backward random process $\overleftarrow{x}(k)$ converges almost surely to a finite limit $x_{\infty}$ if and only if
\begin{equation}\label{eq:ip_Diaco}
\inf_{k>0}\frac{1}{k}\Exp\left[\log\|P(1)\ldots P(k)\|\right]<0.
\end{equation}
If \eqref{eq:ip_Diaco} holds, then the distribution of $x_{\infty}$ is the unique invariant distribution for the Markov process $x(k)$.
\end{lemma}

This result provides conditions for the backward process to converge to a limit. Although the forward process has a different behavior compared to the backward process, the forward and backward processes have the same distribution. This fact allows us to determine, by studying the backward process $\overleftarrow{x}(k)$, whether the sequence of random variables $\{x(k)\}_{k\in\integernonnegative}$ converges in distribution to the invariant distribution of the Markov { {process}} in~\eqref{eq:asynchro-algo}.
%
%
%
%
This analysis is done in the following result.
\begin{lemma}\label{lem:back_convergence}
Consider the random process $x(k)$ defined in~\eqref{eq:asynchro-algo}, where $P(k)$ and $u(k)$ are i.i.d. and have finite first moments $\Exp[P(k)]$ and $\Exp[u(k)]$. If there exists $\alpha\in(0,1]$ such that 
$\Exp[P(k)]=(1-\alpha)I+\alpha P$ where $P$ is Schur stable, then, 
$\overleftarrow{x}(k)$ converges almost surely to a finite limit $x_{\infty}$, and the distribution of $x_{\infty}$ is the unique invariant distribution for $x(k)$.
\end{lemma}
\begin{IEEEproof}
In order to apply Lemma~\ref{Diaco}, let us compute
\begin{align*}
&\inf_{k\in\N}\frac{1}{k}\Exp\left[\log\|\overleftarrow{P}(0,k-1)\|_{1 }\right]\\
&\qquad\leq \inf_{k\in\N}\frac{1}{k}\log\Exp\left[\|\overleftarrow{P}(0,k-1)\|_{1}\right]\\
& \qquad = \inf_{k\in\N}\frac{1}{k}\log\Exp\left[\max_{j\in \V}\sum_{i\in \V}(\overleftarrow{P}(0,k-1))_{ij}\right]\\
& \qquad\leq\inf_{k\in\N}\frac{1}{k}\log\Exp\left[\sum_{j\in \V}\sum_{i\in \V}(\overleftarrow{P}(0,k-1))_{ij}\right]\\
& \qquad\leq\inf_{k\in\N}\frac{1}{k}\log\sum_{j\in \V}\sum_{i\in \V}\Exp\left[\overleftarrow{P}(0,k-1)_{ij}\right]\\
& \qquad\leq\inf_{k\in\N}\frac{1}{k}\log\left(n\left\|\Exp\left[\overleftarrow{P}(0,k-1)\right]\right\|_{\infty}\right)\\
& \qquad =\inf_{k\in\N}\frac{1}{k}\log\left(n\left\|\prod_{h=0}^{k-1}\Exp\left[{P}(h)\right]\right\|_{\infty}\right).
\end{align*}
Let $q$ be the number of distinct eigenvalues of $\Exp[P(k)]$, denoted as 
$\{\lambda_{\ell}\}_{\ell=1}^{q}$, and consider the Jordan canonical decomposition $
\Exp\left[P(k)\right]=UJU^{-1}
$. Then $\left\|\prod_{h=0}^{k-1}\Exp\left[{P}(h)\right]\right\|_{\infty}\leq \|U\|_{\infty}\|J^k\|_{\infty}\|U^{-1}\|_{\infty}$. 
Since the $k$-th power of the Jordan block of size $s$ is 
\begin{align*}&\begin{bmatrix}\lambda & 1 & 0 & \cdots & 0\\
0 & \lambda & 1 & \cdots & 0\\
\vdots & & \ddots & & \vdots\\
0 & \cdots & 0 & \lambda & 1\\
0 & 0 & \cdots  & & \lambda
 \end{bmatrix}^k\\
&\qquad=
\begin{bmatrix}\lambda^k & {k \choose 1} \lambda^{k-1} & {k \choose 2} \lambda^{k-2} & \cdots  & {k \choose s-1} \lambda^{k-s+1}\\ 
0 & \lambda^k & {k \choose 1} \lambda^{k-1}& \cdots  & {k \choose s-2} \lambda^{k-s+2}\\
\vdots & & \ddots & & \vdots\\
0 & \cdots & 0 & \lambda^k & {k \choose 1} \lambda^{k-1}\\
0 & 0 & \cdots  & & \lambda^k
\end{bmatrix},\end{align*}
we deduce that 
$$\|J^k\|_{\infty }=\max_{i\in \V}\sum_{j\in \V}(J^k)_{ij}= \max_{\ell=1, \dots, q}\sum_{m=0}^{s_\ell-1}\lambda_\ell^{k-m}{k\choose m},$$
where $s_\ell$ is the size of the largest  Jordan block corresponding to $\lambda_\ell.$
Then
\begin{align*}\|J^k\|_{\infty }&\leq \max_{\ell=1, \dots, q}|\lambda_\ell|^k\sum_{m=0}^{s_\ell-1}|\lambda_\ell|^{-m}{k\choose m}\\
&\leq\max_{\ell=1, \dots, q}|\lambda_\ell|^kk^{n}\sum_{m=0}^{s_\ell-1}|\lambda_\ell|^{-m}\\
&\leq \chi\rho^kk^{n},\end{align*}
where $\chi$ is a constant independent of $k$ and $\rho$ is the spectral radius of $\Exp[P(k)]=(1-\alpha)I+\alpha P$, which is known to be smaller than 1 because $P$ is Schur stable.
We conclude that there exists a constant $C=\|U\|_\infty \|U^{-1}\|_\infty \chi$, independent of $k$, such that 
$\Exp\left[\log\|\overleftarrow{P}(0,k-1)\|_{1}\right]\leq\log\left(nC\rho^kk^{n}\right)
$
and, consequently,
\begin{align}
\nonumber
&\inf_{k\in\N}\frac{1}{k}\Exp\left[\log\|\overleftarrow{P}(0,k-1)\|_{1}\right]\\
\label{eq:bound-for-Diaco}
&\qquad\leq\lim_{k\to\infty}\frac{\log(C n k^{n}\rho^k)}{k}\\
\nonumber&\qquad=\log\rho<0.
\end{align}
The claim then follows from Lemma~\ref{Diaco}.
\end{IEEEproof}

As a consequence, we deduce that also the (forward) random process $x(k)$ converges in distribution to a limit $x_{\infty}$, and the distribution of $x_{\infty}$ is the unique invariant distribution for $x(k)$.
We are now ready to verify the ergodicity of $x(k)$ under the assumptions of Theorem~\ref{thm:ergodic}.
Let $z(0)$ be a random vector independent from $x(0)$ with the same distribution as $x_{\infty}$. Let $\{z(k)\}_{k\in \integernonnegative}$ be the sequence such that
$$
z(k)=\overrightarrow{P}(0,k-1)  z(0)+\sum_{0\leq \ell\leq k-1}\overrightarrow{P}(\ell+1,k-1)u(\ell)
$$
where $\overrightarrow{P}(\ell+1,k-1)$ is defined as in \eqref{P}.
Since the process $z (k)$ is stationary and the invariant measure is unique we can apply the Birkhoff Ergodic Theorem (see for instance~\cite[Chapter~6]{DS:1993} or \cite[Chapter~5]{MH:06} for a simpler introduction) and conclude that with probability one
$$\lim_{k\to\infty}\frac{1}{k}\sum_{\ell=0}^{k-1}z(\ell)=\Exp[x_\infty].$$
On the other hand, we can compute 
\begin{align}
&\mathbb{P}\left(\|x(k)-z(k)\|_1\geq \eps^k\right)\nonumber\\
&\qquad\leq\frac{\Exp\left[\|\overrightarrow{P}(0,k-1)  (z(0)-x(0))\|_1\right]}{\eps^k} \nonumber\\
&\qquad\leq \frac{\Exp\left[\|\overrightarrow{P}(0,k-1)\|_1\|z(0)-x(0))\|_1\right]}{\eps^k} \nonumber\\
& \qquad\leq \frac{\Exp\left[\|\overrightarrow{P}(0,k-1)\|_1\right]\Exp\left[\|z(0)-x(0)\|_1\right]}{\eps^k} \nonumber\\
& \qquad\leq \frac{Cnk^{n}\rho^{k}}{\eps^k} \Exp\left[\|z(0)-x(0)\|_1\right],\label{eq:conv_distr}
\end{align}
where we have used~\eqref{eq:bound-for-Diaco}.
If we choose $\eps\in(\rho,1)$, then the Borel-Cantelli Lemma \cite[Theorem~1.4.2]{VB:95} implies that with probability one $\|x(k)-z(k)\|_1<\eps^k$ for all but finitely many values
of $k$. Therefore, almost surely $\frac{1}{k}\sum_{\ell=0}^{k-1}\|x(\ell)-z(\ell)\|_1$ converges
 to zero as $k$ goes to infinity, and 
$$\lim_{k\to\infty}\frac{1}{k}\sum_{\ell=0}^{k-1}x(\ell)=\Exp[x_\infty].$$
To complete the proof, we only need to observe that $\Exp[x_\infty]=\lim_{k\to+\infty}\Exp[x(k)]$, which is equal to $x^\star$ as argued after the statement of Theorem~\ref{thm:ergodic}.

\section{Proof of Corollary~\ref{cor:ergodic2}}\label{sec:proofB}
The argument follows the lines of~\cite[Theorem 4.1]{KY-SS-LQ:14}.
Let us define for all $i$ and $k$ in $\integernonnegative$
$$\xi_{ki}=\begin{cases}\omega(i)/\sum_{\ell=0}^{k-1}\omega(\ell)&\text{if}\ i\leq k\\
0&\text{if }i>k.\end{cases}$$
Since $\lim_{k\rightarrow+\infty}\sum_{\ell=0}^{k-1}\omega(\ell)=+\infty$ almost surely, $\{\xi_{ki}\}_{k,i\in\integernonnegative}$ forms a Toeplitz array with probability one.
Since by~\eqref{eq:conv_distr} $\lim_{k\rightarrow+\infty}\|x(k)-z(k)\|_1=0$, we can apply Silverman-Toeplitz Theorem~\cite{OT:11} to conclude that almost surely 
\begin{align*}
&\lim_{k\rightarrow+\infty}\sum_{i=0}^{+\infty}\xi_{ki}\|x(i)-z(i)\|_1\\
&\qquad=\lim_{k\rightarrow+\infty}\frac{1}{\sum_{\ell=0}^{k-1}\omega(\ell)}\sum_{i=0}^{k-1}\omega(i)\|x(i)-z(i)\|_1=0. 
\end{align*}
This equality implies that almost surely
\begin{align*}
\lim_{k\rightarrow+\infty}\frac{1}{\sum_{\ell=0}^{k-1}\omega(\ell)}&\sum_{i=0}^{k-1}\omega(i) x(i)\\
&=
\lim_{k\rightarrow+\infty}\frac{1}{\sum_{\ell=0}^{k-1}\omega(\ell)}\sum_{i=0}^{k-1}\omega(i) (x(i)-z(i))\\
&\qquad+\lim_{k\rightarrow+\infty}\frac{k}{\sum_{\ell=0}^{k-1}\omega(\ell)}\frac{1}{k}\sum_{i=0}^{k-1}\omega(i) z(i)\\
& = \frac{1}{\Exp[\omega(0)]} \lim_{k\rightarrow+\infty} \frac{1}{k}\sum_{i=0}^{k-1} \omega(i) z(i),
\end{align*}
where the last equality comes from the law of large numbers. 
Again by Birkhoff Ergodic Theorem, $\{(\omega(k)^{\top},z(k)^{\top})^{\top}\}_{k\in\integernonnegative}$ is a stationary and ergodic process and we finally conclude
\begin{align*}
&\lim_{k\rightarrow+\infty}\frac{1}{\sum_{\ell=0}^{k-1}\omega(\ell)}\sum_{i=0}^{k-1}\omega(i) x(i)=\frac{1}{\Exp[\omega(0)]}\Exp{[\omega(0)z(0)]}
=x^{\star},
\end{align*}
thanks to the independence between $\omega(0)$ and $z(0)$.

\bibliographystyle{ieeetran}
\bibliography{aliases,biblio-ergodic}
\end{document}